\documentstyle[12pt,graphicx]{article}

\newcommand{\beq}{\begin{equation}}
\newcommand{\eeq}{\end{equation}}
\def\bea{\begin{eqnarray}}
\def\eea{\end{eqnarray}}
\def\nn{\nonumber}

\begin{document}

\newcommand{\sheptitle}{ What can we learn from $\phi_1$ 
                         and $B^0_d \rightarrow 
                         \pi^+ \pi^-$ ?}
\newcommand{\shepauthor}{Tadashi Yoshikawa}
\newcommand{\shepaddressKEK}{
            Theory Group, KEK,
                       Tsukuba, 305-0801, Japan.}

\date{\today}

\begin{titlepage}
\begin{flushright}
hep-ph/0304038\\
KEK-TH-878\\
\today
\end{flushright}
\vspace{.1in}
\begin{center}
{\large{\bf \sheptitle}}
\bigskip \medskip \\
\shepauthor \\
\mbox{} \\
\medskip
{\it \shepaddressKEK} \\
\vspace{.5in}

\bigskip \end{center} \setcounter{page}{0}
\begin{abstract}
We discuss what we can understand from $\phi_1$ and 
$B^0_d\rightarrow \pi^+ \pi^- $ decay mode. 
Using a convention without weak phases 
$\phi_2$ and $\phi_3$, we can solve the parameters from the time-depended 
CP asymmetry. If we can put a condition the contribution from penguin except
for the CKM factor including in the diagram is small, then we can lead the 
allowed region of $R_t$ or $\phi_2$ by using the convention.     
\end{abstract}

\vspace*{6cm}

\begin{flushleft}
\hspace*{0.9cm} \begin{tabular}{l} \\ \hline {\small Email:
tadashi.yoshikawa@kek.jp} \\
\end{tabular}
\end{flushleft}

\end{titlepage}
Measurements of CP phase $\phi_1$ by the Belle\cite{BF1} 
and BaBar\cite{BF2} collaborations
established CP violation in the $B$ meson system.  Measuring 
the other CP phase $\phi_2 $ and $\phi_3$ is also very important to test the 
Kobayashi-Maskawa(KM) model\cite{CKM}. 
The conventional method of measuring $\phi_2$ 
uses the time dependent CP asymmetry in $B^0 \rightarrow \pi^+ \pi^- $
\cite{CONV1, CONV2, LU}. 
However this method has a difficulty of penguin contamination. 
If the contribution from penguin diagram is negligible, the CP asymmetry 
is very clean measurement to extract $\sin 2 \phi_2 $.  But recent
measurement by the Belle\cite{CPVPPBELLE} 
showed the penguin contribution is likely to be 
sizable so that we must take account of them.  
In this letter, we discuss what we can learn about the CP phase $\phi_2$ 
from the present measurements. 
The data we can use in here are the branching ratio $B_{\pi \pi }$,
the coefficients of $\cos\Delta mt$ 
and $\sin\Delta mt $ in the time dependent CP asymmetry, 
$S_{\pi\pi}$ and $A_{\pi\pi}$, and  the weak phase $\phi_1$ measured by 
$B\rightarrow J/\psi K_S $ .     
    
There are two contributions in $B \rightarrow \pi^+ \pi^- $ decay, 
which comes from 
tree and penguin diagrams. The amplitude is 
\bea
A \equiv A(B^0\rightarrow\pi^+\pi^-) = - ( \{T + P_u\}V_{ub}^*V_{ud} 
                                  + P_c V_{cb}^*V_{cd} 
                                  + P_t V_{tb}^*V_{td}  ), 
\eea 
where $T$ is a tree amplitude and $P_i$$(i=u,c,t)$ are penguin amplitudes. 
Using unitarity relation of 
Cabbibo-Kobayashi-Maskawa(CKM) matrix, we can rewrite it to the following 
three type (This is called CKM ambiguity in \cite{LSS}.), 

\noindent
{\bf Convention A: }\\
\bea
A^A =  - ( \{T + P_u - P_t\}V_{ub}^*V_{ud} - \{P_t  - P_c\} V_{cb}^*V_{cd} ) 
  \equiv - ( T_{ut} V_{ub}^*V_{ud} - P_{tc} V_{cb}^*V_{cd} ) 
\eea 
\noindent
{\bf Convention B: }\\
\bea
A^B =  - ( \{T + P_u - P_c\}V_{ub}^*V_{ud} + \{P_t  - P_c\} V_{tb}^*V_{td} )
  \equiv  - (  T_{uc} V_{ub}^*V_{ud} + P_{tc} V_{tb}^*V_{td} ) 
\eea

\noindent
{\bf Convention C: }\\
\bea
A^C =  - ( \{P_t - T - P_u \}V_{tb}^*V_{td} 
            + \{P_c  - T - P_u\} V_{cb}^*V_{cd} ) 
  \equiv  - ( - T_{ut} V_{tb}^*V_{td} - T_{uc} V_{cb}^*V_{cd} )  
\eea
Using the convention A and B\cite{CONV1, CONV2} 
one can extract a phase $\phi_2$ from 
time dependent CP asymmetry if the penguin contribution is negligible.  
However present situation is not so. We can not extract the value 
because the unknown parameters are too many\cite{LSS}. 
So we consider how to use the remaining case C.  
Convention C is including  
only $\phi_1$ which is found from $B\rightarrow J/\psi K_s$ mode et.al.  
Hence we can find all parameters by using $\phi_1$ and 
the measurements of CP asymmetry .  

We rewrite the decay amplitude for each Convention as follows: 
\bea 
A^A &\equiv& - |T_{ut}| |V_{ub}^*V_{ud}| e^{i\delta_{Tut}} 
              \left( e^{i\phi_3}+r_A e^{i\delta^A} \right), \\
A^B &\equiv& - |T_{uc}| |V_{ub}^*V_{ud}| e^{i\delta_{Tuc}} 
              \left( e^{i\phi_3}+r_B e^{i\delta^B} e^{-i\phi_1} \right), \\
A^C &\equiv&  - |T_{uc}| |V_{cb}^*V_{cd}| e^{i\delta_{Tuc}} 
           \left( 1-r_C e^{i\delta^C} e^{-i\phi_1} \right), 
\eea
where $\delta $ is the strong phase and 
\bea
r_A = \frac{|P_{tc}||V_{cb}^*V_{cd}|}{|T_{ut}||V_{ub}^*V_{ud}|},~~~
r_B = \frac{|P_{tc}||V_{tb}^*V_{td}|}{|T_{uc}||V_{ub}^*V_{ud}|},~~~
r_C = \frac{|T_{ut}||V_{tb}^*V_{td}|}{|T_{uc}||V_{cb}^*V_{cd}|}. 
\eea 
We can find the relation among the conventions from $A^A=A^B=A^C$ or 
$ T_{uc} - T_{ut} = P_{tc} $ as following, 
\bea
R_t - r_C \cos\delta^C &=& r_B R_b \cos\delta^B,  
\label{rccos} \\
-r_C \sin\delta^C &=& r_B R_b \sin\delta^B,
\label{rcsin} 
\eea
where 
\bea
R_b &=& \frac{|V_{ub}^*V_{ud}|}{|V_{cb}^*V_{cd}|} 
        = \frac{\sin\phi_1}{\sin\phi_2}, \\
R_t &=& \frac{|V_{tb}^*V_{td}|}{|V_{cb}^*V_{cd}|} 
        = \frac{\sin\phi_3}{\sin\phi_2} = R_b \cos\phi_2 + \cos\phi_1 ,
\eea
and the relations among the parameters for each conventions are
\bea
r_B &=& r_A ~ r_C \\
\delta^A &=& \delta^B - \delta^C 
\eea

The measurements for $B\rightarrow\pi^+\pi^-$ are 
\bea
\Gamma(B^0\rightarrow \pi^+ \pi^-) + \Gamma(\bar{B}^0\rightarrow \pi^+ \pi^-) 
                             &\propto & (|A|^2 + |\bar{A}|^2 ) \\
\Gamma(B^0\rightarrow \pi^+ \pi^-) - \Gamma(\bar{B}^0\rightarrow \pi^+ \pi^-) 
         &\propto &
         (|A|^2 - |\bar{A}|^2 ) \cos\Delta m t  \nn \\
            & &  - 2 Im(e^{-2i\phi_1} A^* \bar{A}) \sin\Delta m t .  
\eea
The correspondence to the measurements in convention C are  
\bea
B_{\pi \pi} &\propto& (|A|^2 + |\bar{A}|^2 ) 
            = 2 |T_{uc}|^2 | V_{cb}^*V_{cd}|^2 
              \{ 1 + r_C^2 - 2 r_C \cos\delta^C \cos\phi_1 \} 
\label{BppofC}\\
A_{\pi \pi} &\equiv& - \frac{|A|^2 - |\bar{A}|^2 }{|A|^2 + |\bar{A}|^2}
   = \frac{2 r_C \sin\delta^C \sin\phi_1 }
          { 1 + r_C^2 - 2 r_C \cos\delta^C \cos\phi_1 }
\label{AppC}\\
S_{\pi \pi} &\equiv& 
            \frac{2 Im(e^{-2i\phi_1} A^* \bar{A})}{|A|^2 + |\bar{A}|^2} 
            = \frac{ - \sin2\phi_1 
              + 2 r_C \cos\delta^C \sin\phi_1}
           { 1 + r_C^2 - 2 r_C \cos\delta^C \cos\phi_1 }
\label{SppC}
\eea 
From the three measurements we can find the three parameters 
$|T_{uc}|$, $r_C$ and $\delta^C$ by inputting the value of $\phi_1$ 
as the world average which is  
$\phi_1=23.6^\circ \pm 2.4^\circ$\cite{NIR}.  

Recently, Belle collaboration reported the results\cite{CPVPPBELLE} 
but they are not still 
consistent with the values by BaBar\cite{CPVPPBABAR}. Hence, we discuss 
by using the average between Belle and BaBar. The data are in Table 1.  

The first, we find the allowed region for $r_C$ and $\delta^C$ from 
eqs. (\ref{AppC}) and (\ref{SppC}) by inputting 
$\phi_1 = 23.6 ^\circ\pm 2.4^\circ$, 
$S_{\pi \pi} = -0.47 \pm 0.26 $ and $ A_{\pi \pi} = 0.51 \pm 0.19 $ 
and it is shown in Fig. 1.  The solution 
for the central value are 
\bea 
(r_C,\delta^C) = ( 0.816, 8.6^\circ )~~\mbox{and}~~ ( 0.770, 71.6^\circ ).
\eea   
We can find that the solutions have a discrete ambiguity and 
there are two regions in Fig.1. 
One of them is smaller region near $\cos\delta^C \sim 1 $ and around 
$r_C \sim 0.8 $ and this show the case of small penguin contribution. 
$\delta^C$ is the angle between $T-P_u-P_t$ and $T-P_u-P_c$ and 
it comes from the difference between $P_t$ and $P_c$. If top-penguin is 
very close to charm-penguin or the tree contribution $T$ dominant, 
\bea
\frac{ T-P_u-P_t}{ T-P_u-P_c} \sim  1 - \frac{P_t-P_c}{T} \sim 1, \nn 
\eea
and then $\delta^C$ becomes to small angle and $r_C$ becomes to close $R_t$. 
We can guess this region is reasonable. 

In the Convention C, we can solve and find the all parameters.  However, 
because the the main target in this mode is to extract $\phi_2$, we have to 
consider also the other convention or to convert the solution in convention C
to the others.  

\begin{table}
\begin{center}
\begin{tabular}{|c|c|c|c|}\hline 
  & Belle\cite{CPVPPBELLE} & BaBar\cite{CPVPPBABAR} & Average \\
\hline
$Br(B\rightarrow \pi\pi) \times 10^{5} $ & 0.54 $\pm$ 0.12 $\pm$ 0.05 
                          & 0.47 $\pm$ 0.06 $\pm$ 0.02
                          & 0.48 $\pm$ 0.05  \\
\hline
$S_{\pi \pi}$ & -1.23 $\pm$ 0.41 $^{+0.08}_{-0.07}$ 
        & 0.02 $\pm$ 0.34 $\pm$ 0.05 & -0.47 $\pm$ 0.26 \\
$A_{\pi \pi}$ & 0.77 $\pm$ 0.27 $\pm$ 0.08 & 0.30 $\pm$ 0.25 $\pm$ 0.04 
        &   0.51 $\pm$ 0.19 \\ 

\hline
\end{tabular}
\caption{The experimental data and the average. }
\end{center}
\end{table}

\begin{figure}[h] 
\begin{center}
\includegraphics[scale=0.48,angle=-90]{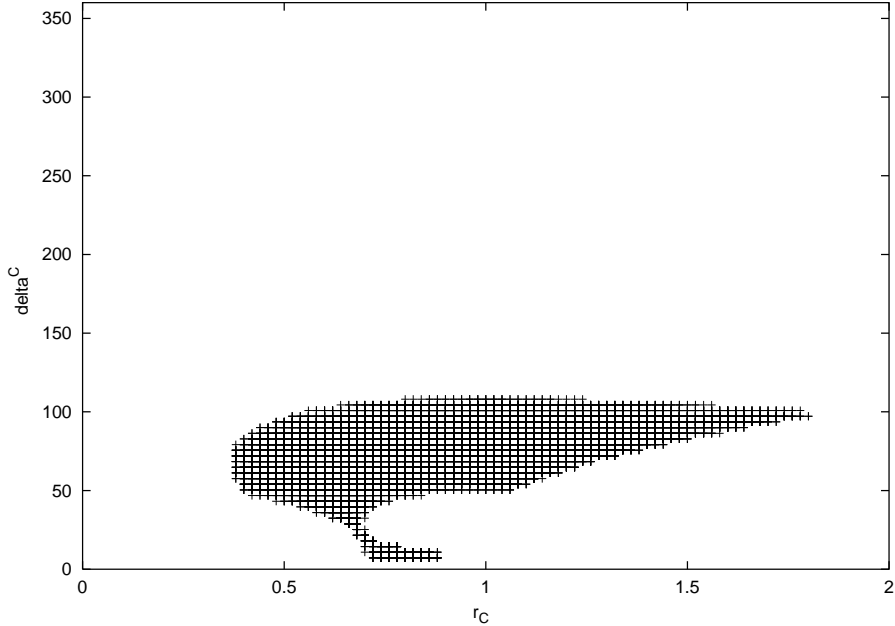} 
\caption{The allowed region for the $r_C$ and $\delta^C $ for 
$S_{\pi\pi}=-0.47\pm0.26, A_{\pi\pi}=0.51\pm0.19$ and 
$\phi_1 = 23.6^\circ \pm 2.4^\circ $. }
\end{center}
\label{fig:RCvsdeltaC}
\end{figure}

Unfortunately, we can not convert from the solutions in Convention C to 
the other convention case\cite{LSS}, 
because the relation to extract them are just only 
2 equations (\ref{rccos}) and (\ref{rcsin}) 
for 3 parameters, $r_B,\delta^B $ and $\phi_2$. 
So we consider a reasonable situation that $r_A$ and $r_B$ are not so small 
but,indeed, it was enhanced by $1/R_b$ and the ratio between tree and 
penguin without KM factor is small. 
We guess the ratio is order of 0.1.  If so, 
\bea
r_C &=& \frac{|T_{uc}-P_{tc}|}{|T_{uc}|} R_t \\
    &\sim& R_t - r_B R_b \cos\delta^B 
   + \frac{1}{2}r_B^2\frac{R_b^2}{R_t}
             \left(1-\frac{1}{2}\cos^2\delta^B\right) + O(\{P/T\}^3)
\eea 
and the higher order of $P/T$ without KM factor is neglected. 
Using the relation (\ref{rccos}) and (\ref{rcsin}), 
we find a equation for $R_t$ or $\phi_2 $, 
\bea
R_t^2 + 2 R_t r_C ( \cos\delta^C - 2 ) + r_c^2 ( 2 - \cos^2\delta^C ) = 0 
\label{rteq}
\eea 
From this equation, we find the allowed region on $r_C-\phi_2$ plane for 
the region we found from the averaged experimental values about 
$B\rightarrow\pi \pi$ decay mode.  The results for the central values are 
shown in Table 2. If the magnitude of penguin amplitude is negligible, then
the $R_t$ should be very close to $r_c$. 
In $r_C = R_t $ case, the dependence of $r_c$ for $\phi_2$ 
is shown as the solid line for 
$\phi_1= 23.6^\circ $ and dashed lines for $\phi_1= 21.2^\circ $ 
and $26.0^\circ $ in Figs.2-4. 
The close region to the line of $r_C = R_t $ will 
satisfy the small penguin condition. To check this, we show $r_B$ for $R_t$ 
we obtained in Table 2.  For $(r_C, \delta^C)=(0.82,8^\circ)$, 
$r_B$ is about $0.3$ and the ratio between penguin and tree without KM factor 
$|P_{tc}/T_{uc}| (\sim r_B R_b )$ is about 0.1.  
For $(0.77,72^\circ)$, $r_B$ is larger 
than 1 and this case is out of the assumption. 

\begin{table}[h] 
\begin{center}
\begin{tabular}{|c|c|c|c||c|}\hline 
 $r_C$ & $\delta^C $ & $R_t$ & $\phi_2$ & $r_B$ \\
\hline
0.816 & 8.4$^\circ $ & 0.812 & 104.6$^\circ $ & 0.287\\
      &       & 0.837 & 101.3$^\circ$ & 0.300\\
0.770 & 71.6$^\circ $ & 0.552 & 132.3$^\circ$ & 1.46 \\
      &       & 2.041 & 19.6$^\circ$ & 1.63 \\
\hline
\end{tabular}
\caption{ $R_t$ and $\phi_2$ for the solution of $r_C$ and $\delta^C $ for 
          the central values of experimental data. }
\end{center}
\end{table}

If the magnitude of penguin contributions are very small, $r_C = R_t$ and 
we can approximately extract the value of $R_t$. Even if it is not so small, 
we can guess $r_c$ should be near to $R_t$ and 
the cosine of the strong phase $\delta^C$ should be close to 1. It 
is in the smaller region near $360^\circ $ in Fig.1. If we take account of 
only region for $\delta^C$ less than $30^\circ $ and $10^\circ $, then 
the region for $\phi_2$ are reduced and remain only region 
around $\phi_2 = 100^\circ $. See Fig.3.  
When we consider the case the experimental error of 
$S_{\pi\pi}$ and $A_{\pi\pi}$ are reduced up to $0.1$, the allowed region 
will become smaller. We show the regions in Fig.4. 
The region for $r_c$ and $\delta^C$ completely separate to two parts. 
The region near $0^\circ $ show that $\phi_2$ is around $100^\circ$.   

\begin{figure}[htbp] \hspace*{40mm}
\includegraphics[scale=0.3,angle=-90]{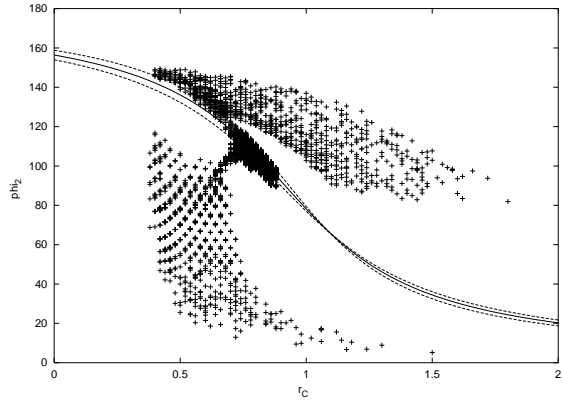} 
    \label{fig:RCvsphi2}
\caption{The allowed region of $\phi_2$ obtaining from $r_C$ and $\delta^C$ 
         in Fig.1 by using eq.(\ref{rteq}).  The lines show $r_C=R_t$ for 
         $\phi_1= 23.6^\circ $(solid line) and for $\phi_1= 21.2^\circ $ 
         and $26.0^\circ$ (dashed lines) } 
\end{figure}

\begin{figure}[htbp]
\begin{center}
\begin{minipage}[c]{0.4\textwidth}
\includegraphics[scale=0.27,angle=-90]{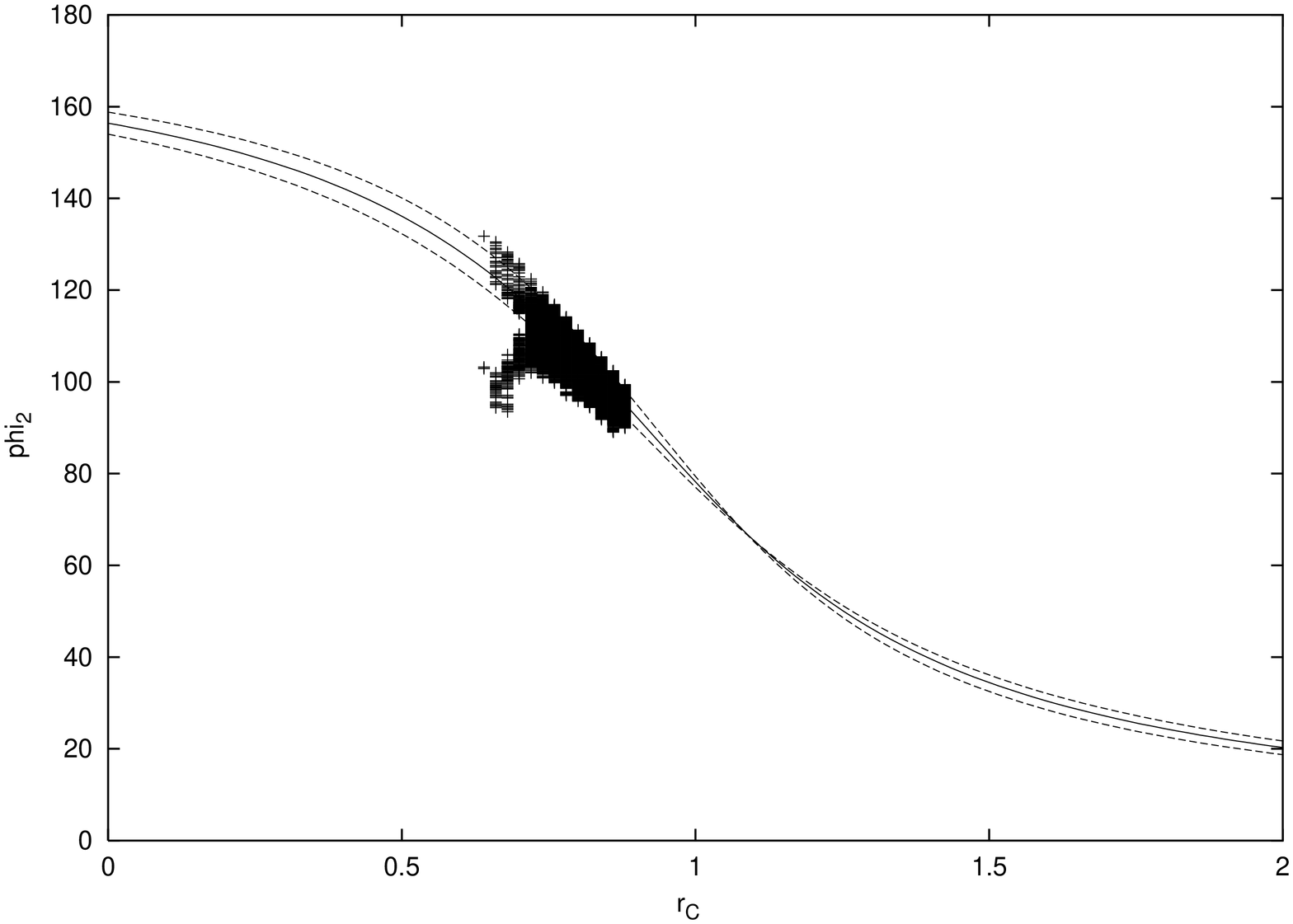} 
\end{minipage}
    \hspace*{5mm}
\begin{minipage}[c]{0.4\textwidth}
\includegraphics[scale=0.27,angle=-90]{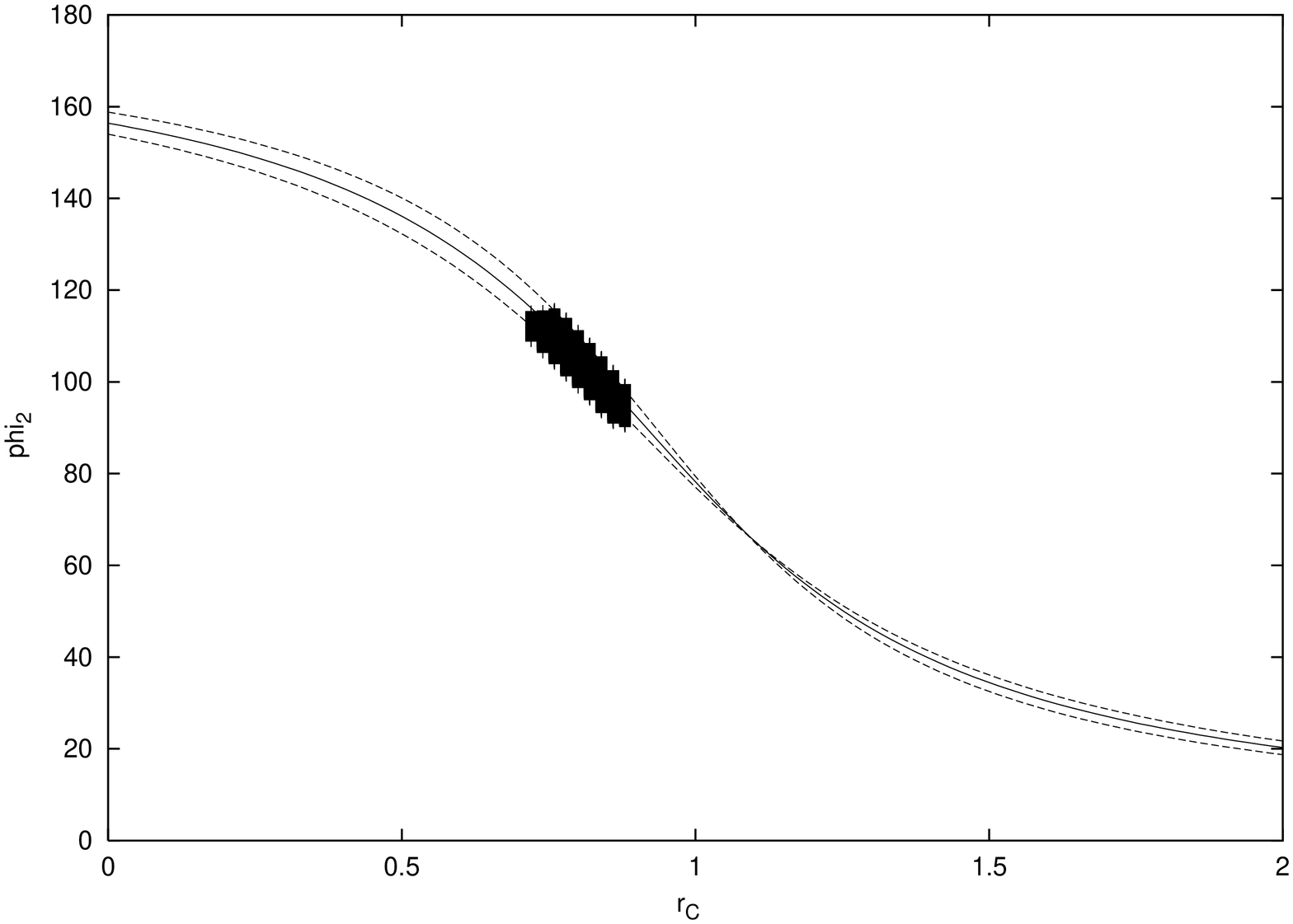} 
\end{minipage}
\caption{$\phi_2$ when we cut the region of $\delta^C$ 
over $30^\circ$ (left) and $10^\circ $ (right). } 
    \label{fig:RBvsphi2OVER350}
\end{center}
\end{figure}

\begin{figure}[htbp]
\begin{center}
\begin{minipage}[c]{0.4\textwidth}
\includegraphics[scale=0.27,angle=-90]{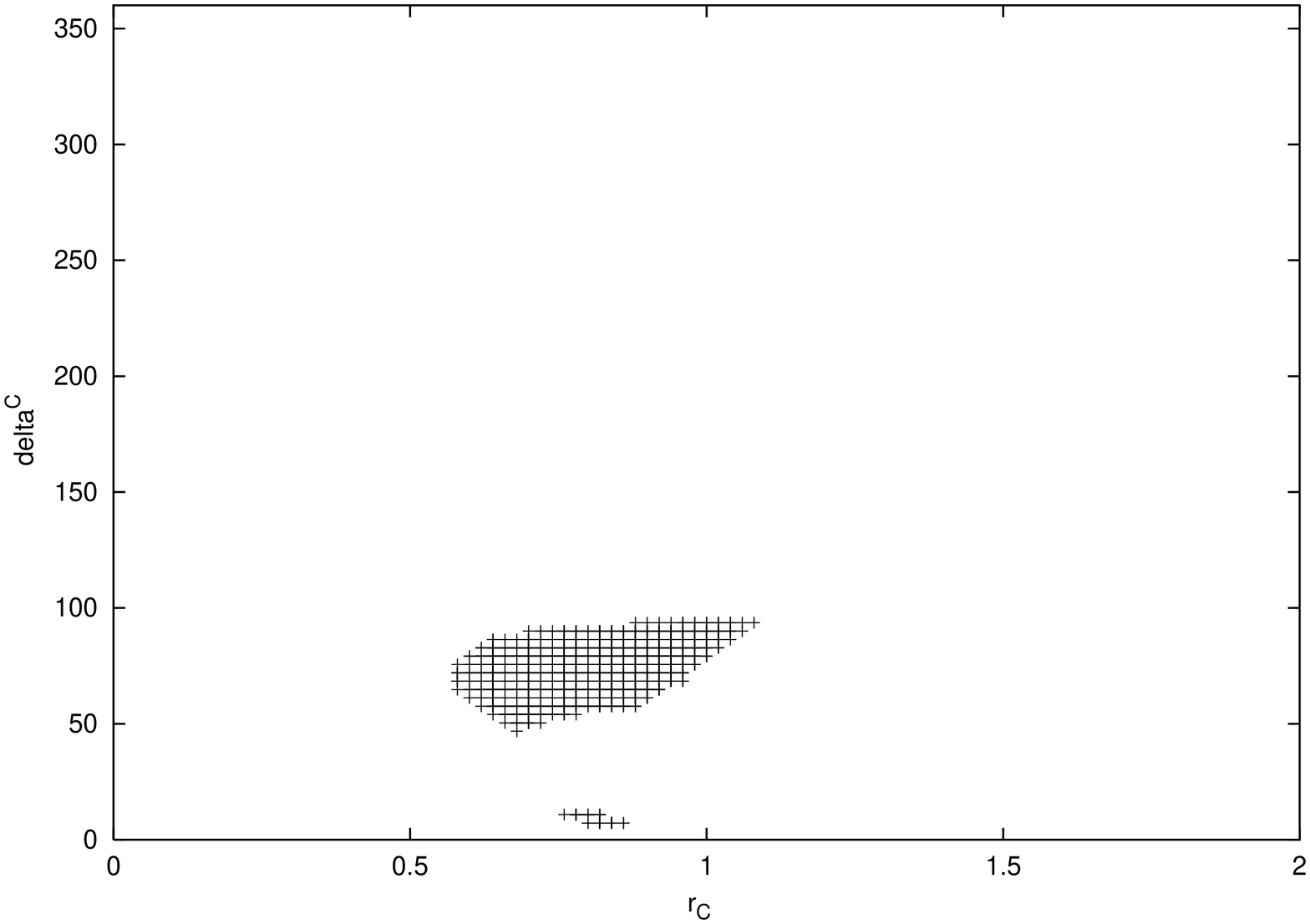} 
\end{minipage}
    \hspace*{5mm}
\begin{minipage}[c]{0.4\textwidth}
\includegraphics[scale=0.27,angle=-90]{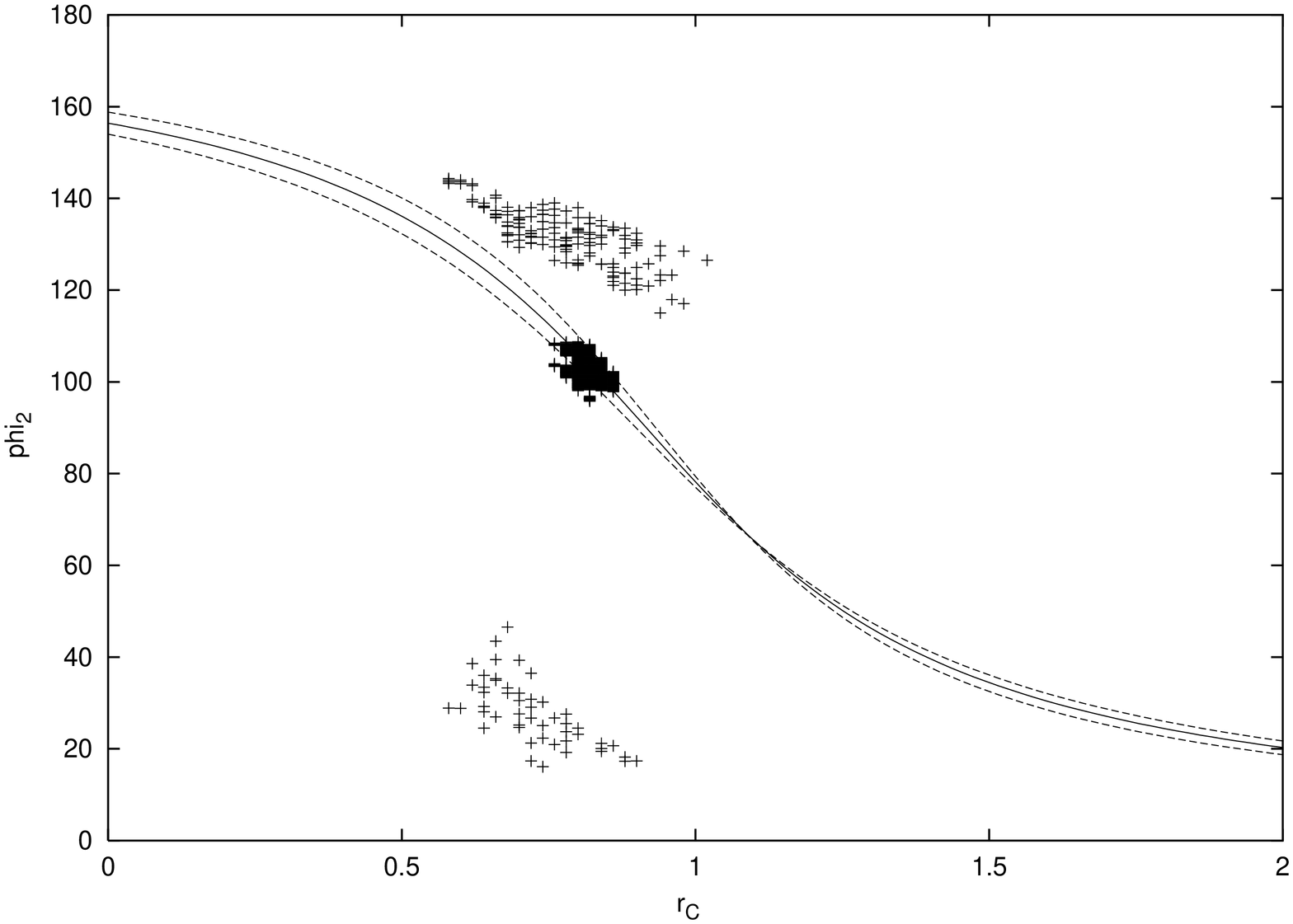} 
\end{minipage}
\caption{The allowed region of $r_C, \delta^C$ and $\phi_2$ in the case the 
error of $A_{\pi\pi}$ and $S_{\pi\pi }$ is less than 0.1. }
    \label{fig:RBvsphi2ER0.1}
\end{center}
\end{figure}

In this letter we discussed how to use the solution in Convention C. 
Though the experimental values of $A_{\pi\pi}$ and $S_{\pi\pi} $ are not 
still consistent between Belle and BaBar, we used the averaged value and 
obtained the allowed region within the error.  
If the solution near $\delta^C = 0^\circ$ is true, namely  
the ratio of tree and penguin without KM factor is as small as order 0.1, 
then we can estimate $R_t$ or $\phi_2$ and $\phi_2$ is around $100^\circ $. 

\section*{Acknowledgments}
We would like to thank S. Oh for discussion. The main part of this work 
was done when T.Y. belonged to International Center for Elementary 
Particle Physics(ICEPP) in Tokyo University as a COE research fellow.

\end{document}